\documentclass[oldversion,a4paper]{aa}
\usepackage{graphics}

\newcommand{\kms}{km\,s$^{-1}$} 

\newcommand{\tev}[1]{$T_{\rm eff}= #1$~K}

\newcommand{\lggv}[1]{$\log g= #1$}
\newcommand{\vs}{$v_{\rm e}\sin\,i$}
\newcommand{\vsv}[1]{$v_{\mathrm{e}}\sin i=#1$}

\newcommand{\hd}{HD\,137509}
\newcommand{\bz}{$\langle B_{\rm z} \rangle$}
\newcommand{\bs}{$\langle B \rangle$}
\newcommand{\bx}{$v_{\rm e}\sin\/i\/\langle xB_{\rm z} \rangle$}
\newcommand{\bq}{$\langle B^2_{\rm z} + B^2 \rangle^{1/2}$}

\newcommand{\fifps}[2]{\centering\resizebox{#1}{!}{\includegraphics{#2}}}
\newcommand{\figps}[1]{\resizebox{\hsize}{!}{\rotatebox{0}{\includegraphics{#1}}}}

\begin{document}

\title{Remarkable non-dipolar magnetic field of the Bp star HD\,137509\thanks{Based 
on the data from the UVES Paranal Observatory Project (ESO DDT Program ID 266.D-5655).}} 

\author{O. Kochukhov}

\institute{Department of Astronomy and Space Physics, Uppsala University, SE-751 20, Uppsala, Sweden}

\date{Received 31 January 2006 / Accepted 26 March 2006}

\abstract{The southern magnetic Bp star \hd\ exhibits complex rotational modulation of the longitudinal field
and other magnetic observables. Interpretation of this magnetic variability in the 
framework of the low-order multipolar field models suggests a very strong quadrupolar component to
dominate the surface field topology of \hd. I have examined the high-quality VLT/UVES spectra of \hd\ 
and discovered resolved Zeeman split components in some of the spectral lines.
The inferred mean surface field modulus, $\langle B \rangle=29$~kG, agrees with the multipolar model
predictions. This confirms the presence of an extremely strong non-dipolar
magnetic field in \hd\ and establishes this star as the object with the second-largest field among
magnetic chemically peculiar stars.}


\maketitle

\section{Introduction}
\label{intro}

Significant fraction of the middle and upper main sequence stars of the B--F spectral classes shows
unusual surface properties. The chemical composition of the outer layers in these stars departs
significantly from solar, with elements distributed non-uniformly both horizontally over the
stellar surface and vertically in the line forming region. Many of these \textit{chemically
peculiar} (CP) stars also possess a strong, well-ordered magnetic field. Unlike the rapidly
evolving, complex spotted magnetic topologies of active late-type stars, CP-star magnetic fields
are stable and cover the whole stellar surface. The field strength ranges from a few hundred G 
up to $\approx$\,35~kG. The field is frozen in the stellar plasma and its observed changes are not
intrinsic to the star but rather a result of the rotational modulation of the visible projection
of the stellar magnetic field geometry.

The structure and origin of the global magnetic fields of CP stars is a matter of intense debate.
The early observational magnetic field studies focused on the measurements and establishing the
form of the rotational modulation of the disk-integrated line of sight field component --
longitudinal field \bz. The vast majority of the magnetic CP stars exhibit smooth, single-wave
variation of longitudinal field during the stellar rotation cycle. This was considered as an evidence
for the dominant dipolar contribution to the field topology (e.g. Borra \& Landstreet
\cite{BL80}). However, after several additional magnetic observables were incorporated in the
field modelling procedure (Landstreet \& Mathys \cite{LM00}; Bagnulo et al. \cite{BLL02}), it was
found that purely dipolar fields are often insufficient to explain observations, hence more
complex quadrupolar and octupolar fields had to be introduced. Moreover, recent detailed analysis
of the four Stokes parameter spectra of CP stars (Kochukhov et al. \cite{KBW04}) revealed that in
some stars exceedingly complex magnetic geometries may give rise to a deceptively simple, nearly
sinusoidal \bz\ variation.

Thus, despite a certain success of the phenomenological low-order multipolar field models, it
remains unclear whether the suggested quadrupolar and octupolar deviations from the basic dipolar
structure reflect the physical reality or represent merely a mathematical approximation of the
unresolved small-scale field. Solution of this problem is certainly of great interest because
it will provide additional important constraints for the theoretical models of the field generation
and evolution in CP stars. I suggest that a better understanding of the nature of non-dipolar
fields in early-type magnetic stars can be gained through detailed investigation of stars with the
\textit{dominant} non-dipolar field components. Up to now only two objects with extraordinary
double-wave longitudinal field curves, hence clearly non-dipolar fields, have been found. These
stars are HD\,37776 (Thompson \& Landstreet \cite{TL85}; Khokhlova et al. \cite{KVS00}) and \hd\
(Mathys \& Lanz \cite{ML97}). The third star, HD\,133880  (Landstreet \cite{L90}) shows a strongly
distorted single-wave \bz\ variation and is often referred to as another example of the star with 
an unusually strong non-dipolar field.

One could hope to constrain the structure of non-dipolar fields in these objects and distinguish
a low-order multipolar field from a more complex, spot-like, magnetic topology with the help of
the disk-integrated surface field strength \bs\ measurement. The low-order multipolar field
scenario predicts that a relatively low-amplitude non-sinusoidal longitudinal field variability
is produced by a quadrupolar or octupolar field of an enormous strength. For instance, a
4~kG peak-to-peak modulation of \bz\ in HD\,37776 (Thompson \& Landstreet \cite{TL85}) should be
interpreted as a signature of $>$\,50~kG surface field (Bohlender \cite{B94}). In contrast,
discarding a low-order multipolar field parameterization, one can successfully reproduce the
observed non-sinusoidal \bz\ phase curves with a high-contrast surface field of a modest strength
(only a factor of 1.5--2 larger than the \bz\ extremum). 

In this paper I try to constrain different possible non-dipolar field structures
in the Bp star \hd. Of the three early-type stars with distinctively non-dipolar magnetic field
configurations, \hd\ has the longest rotation period, lowest projected rotational velocity and,
therefore, presents the best opportunity for the direct diagnostic of the surface field strength
through the high-resolution spectroscopic observations and line profile analysis.

The rest of the paper is organized as follows. In Sect.~\ref{atmrot} I summarize the physical
properties of \hd\ and refine the value of the stellar rotation period. Sect.~\ref{mag} overviews
previous magnetic field measurements and presents interpretation of the magnetic field
variability in terms of the low-order multipolar field topologies. Detection of the resolved
Zeeman split lines in the spectrum of \hd\ and the first field strength measurement are reported
in Sect.~\ref{split}. The study concludes with a summary and discussion in Sect.~\ref{discus}.

\section{Atmospheric parameters and rotation period}
\label{atmrot}

\hd\ (HIP 76011, NN~Aps) was assigned a B9p SiCrFe spectral classification by Cowley \& Houk
(\cite{CH75}). These authors mention strong lines of \ion{Si}{ii}, \ion{Cr}{ii}, \ion{Fe}{ii}, 
\ion{Ti}{ii}, and the absence of \ion{He}{i} features. Atmospheric parameters of \hd, \tev{12600\pm200}
and \lggv{4.06\pm0.11}, were determined by Hunger \& Groote (\cite{HG99}) using the IR flux
method and comparison with the stellar evolutionary tracks. On the other hand, Hubrig et al.
(\cite{HNM00}) suggested \tev{11900\pm600} from the published spectral type and deduced
\lggv{3.87\pm0.16} from the Hipparcos parallax and theoretical stellar models. I find that the
H$\beta$ and H$\gamma$ hydrogen lines in the UVES spectrum of \hd\ discussed below
(Sect.~\ref{split}) are best fitted with \tev{12750\pm500} and \lggv{3.8\pm0.1} if the enhanced
metallicity $[M/H]=+1.0$ {\sc atlas9} (Kurucz \cite{K93}) stellar model atmosphere is employed.

Photometric variability of \hd\ was detected by Lod\'en \& Sundman (\cite{LS89}). Mathys
(\cite{M91}) discovered a strong reversing longitudinal magnetic field and pointed to spectacular
variations of the \ion{Si}{ii} and iron-peak element line intensities, confirming earlier tentative
spectrum variability reports by Cowley \& Houk (\cite{CH75}) and Pedersen (\cite{P79}). Mathys \&
Lanz (\cite{ML97}) have combined all available magnetic measurements and obtained new time-series
photometry in the Geneva system. Simultaneous consideration of these observables allowed the
authors to obtain a refined value $P=4.4916\pm0.0002$~d for the rotation period of \hd.
 
\begin{figure}[!t]
\fifps{8cm}{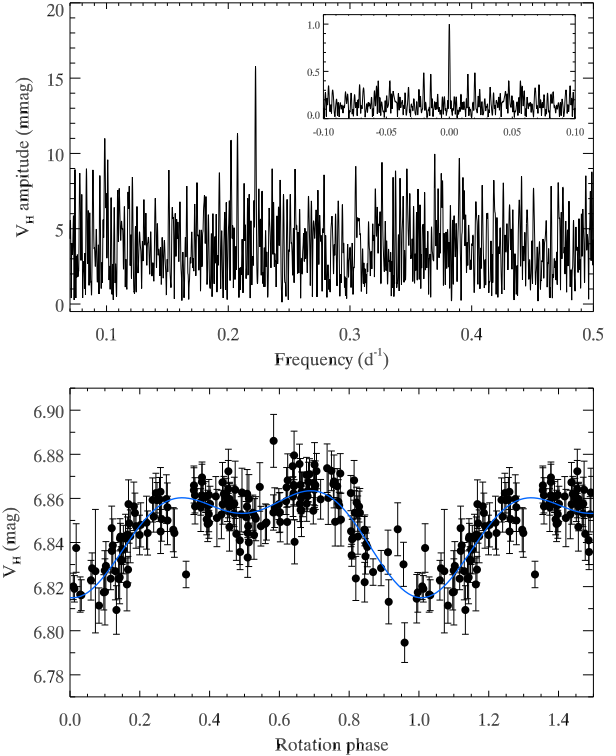}
\caption{Time-series analysis of the Hipparcos epoch photometry of \hd. \textit{Upper panel:} amplitude spectrum 
of the photometric variation. The inset shows the window function of the Hipparcos observations. 
\textit{Lower panel:} photometric measurements folded with the period $P=4.49257$~d.
The solid curve illustrates the least-squares fit by a sinusoid and its first harmonic.}
\label{fig1}
\end{figure}

In addition to the previous reports of periodic brightness modulation, \hd\ also shows pronounced
variability in the Hipparcos epoch photometry (ESA \cite{ESA}). The Hipparcos satellite observed
the star over the period of 3.2 years, obtaining 192 individual photometric measurements. Fourier
analysis of these data is illustrated in Fig.~\ref{fig1}. The only statistically significant peak
is detected at the frequency $\nu=0.223$~d$^{-1}$. The least-squares fit by a cosine function
and its first harmonic 
\begin{equation} 
\begin{array}{rl}
V_H = V_0 ~~+ & V_1 \cos{[2\pi (T-T_0)/P + \varphi_1]}  \\
+ & V_2 \cos{[4\pi (T-T_0)/P + \varphi_2]} 
\end{array}
\end{equation} 
yields $P=4.49257\pm0.00029$~d, $V_0=6.846$~mag, $V_1=0.019\pm0.001$~mag, $V_2=0.012\pm0.001$~mag.
Here the epoch of positive magnetic extremum $JD=2448344.575$ (Lanz \& Mathys \cite{LM91}) was 
adopted as the reference time $T_0$. 

The new estimate of the rotation period differs by $<$\,$3\sigma$ from the value derived by Mathys
\& Lanz (\cite{ML97}) and has similar formal precision. Moreover, the present analysis of the 
Hipparcos photometric time-series overcomes aliasing problem typical of the previous photometric
observations (see fig.~1 in Mathys \& Lanz \cite{ML97}) and hence gives more confidence in the
derived period.

The brightness variation of \hd\ as measured by Hipparcos correlates remarkably well with the
magnetic field modulation. Comparison of the photometric curve (Fig.~\ref{fig1}) and \bz\
variability (Fig.~\ref{fig2}) indicates that the star shows maximum brightness at the phase of
positive magnetic extremum ($\varphi=0.0$). The secondary longitudinal field maximum at
$\varphi=0.5$ also has its counterpart in the photometric curve.

\section{Magnetic field topology}
\label{mag}

The first longitudinal field measurements were obtained for \hd\ by Mathys (\cite{M91}) using metal
line spectropolarimetry. Additional \bz\ observations based on the photopolarimetric measurements in
the H$\beta$ line were acquired by Bohlender et al. (\cite{BLT93}). Variation of other
integral magnetic observables, the crossover \bx\ and the mean quadratic field \bq, was studied 
by Mathys (\cite{M95a,M95b}) and Mathys \& Hubrig (\cite{MH97}).

\begin{figure}
\figps{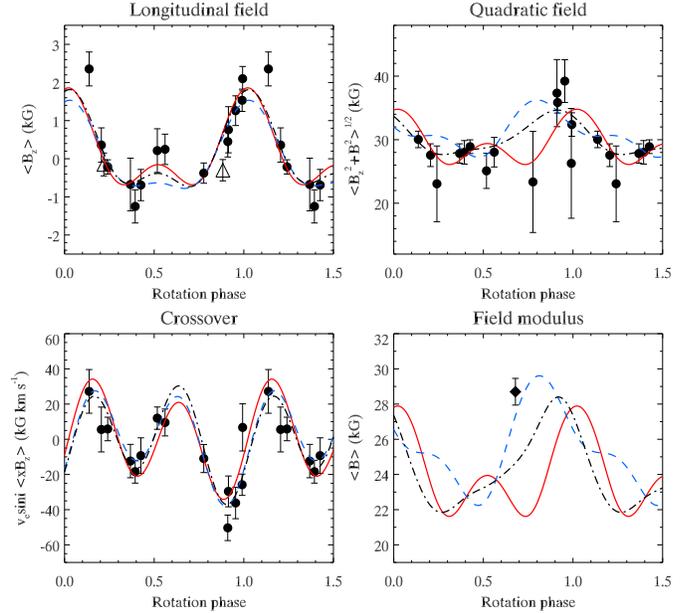}
\caption{Observed variation of the magnetic field moments in \hd\ and predictions
of the best-fit multipolar models of the magnetic field topology. The longitudinal field, 
quadratic field, and crossover observations are taken
from Mathys (\cite{M91,M95a,M95b}), Mathys \& Hubrig (\cite{MH97}) (\textit{filled circles}), 
and Bohlender et al. (\cite{BLT93}) (\textit{open triangles}). A single data point
shown in the lower right panel represents the field modulus measurement reported in the present paper. 
Smooth curves show the fit by multipolar models discussed in the text: 
model $A$ (\textit{solid line}), model $B$ (\textit{dashed line}), and model $C$ (\textit{dash-dotted line}).}
\label{fig2}
\end{figure}

All longitudinal field measurements available for \hd\ are illustrated in Fig.~\ref{fig2}. In total
there are 16 \bz\ determinations, but only four of them correspond to a $>$\,$3\sigma$ field detection.
Nevertheless, the longitudinal field variation of \hd\ is reasonably well defined and exhibits a 
remarkable double-wave shape, with a peak-to-peak amplitude of $\approx$\,3.5~kG and unusually narrow 
positive field extremum at rotation phases $\varphi$\,=\,0.8--1.2. An even more pronounced double-wave
variation is evident in the crossover measurements (see Fig.~\ref{fig2}). The mean quadratic field
reaches up to 36~kG, which is the largest \bq\ found in magnetic CP stars (Mathys \cite{M95b}). These
measurements provide a hint for the presence of a very strong surface field in \hd. However, the
quadratic field diagnostic by itself is rather inconclusive because it relies on the interpretation of
subtle magnetic line broadening effects and neglects horizontal chemical inhomogeneities. For
several spotted magnetic CP stars very large quadratic fields inferred by Mathys (\cite{M95b}) could
not be confirmed by the subsequent detailed line profile analyses of high-resolution spectra (Kupka et al.
\cite{KRW96}; Kochukhov et al. \cite{KDP04}).

Uncertainties of the magnetic observables notwithstanding, I have attempted to probe the field
structure of \hd\ in the framework of the low-order multipolar magnetic geometries and using all
available measurements of \bz, \bx, and \bq. The purpose of this exercise is not to study details of the 
stellar magnetic field topology, but to assess the range of the disk-averaged surface field
strength \bs\ expected for such simple models. The multipolar field modelling procedure applied to \hd\
is similar to the analysis by Landstreet \& Mathys (\cite{LM00}) and Bagnulo et al. (\cite{BLL02}).
Having no \textit{a priori} reason to prefer one or another form of the multipolar field
parameterization suggested in the literature, I have tested the three most widely used types of 
magnetic models. The first one (model $A$) is identical to the field parameterization used by 
Landstreet \& Mathys (\cite{LM00}). The field is represented by an aligned superposition of dipole,
axisymmetric quadrupole and octupole. The 6 free model parameters include the respective field
strengths $B_{\rm d}$, $B_{\rm q}$, $B_{\rm o}$, two angles, $\beta$ and $\gamma$, giving orientation of
the magnetic field axis, and inclination $i$ of the stellar rotation axis. The second model ($B$)
disposes of the field symmetry assumption and approximates the magnetic structure by a sum of dipole
and linear quadrupole, not necessarily aligned with respect to each other (Khokhlova et al.
\cite{KVS00}). This model has 7 free parameters: the field intensities $B_{\rm d}$ and $B_{\rm q}$,
four angles specifying orientation of the dipolar and quadrupolar components, and stellar inclination.
Finally, model $C$ is the one employed by Bagnulo et al. (\cite{BLL02}). It is similar to model $B$,
but includes non-linear quadrupole, whose geometrical properties are now defined with the 4 angles. 
Since the field in \hd\ is dominated by a higher-order magnetic component, it is convenient
to reckon the azimuth angles $\gamma_1$ and $\gamma_2$ from the same zero-phase plane as the dipolar
azimuth angle $\gamma$. The projected rotational velocity, \vs, influences the amplitude of
the \bx\ variation and is adjusted for each of the three magnetic models to fit the crossover observations.

\begin{table}[!t]
\caption{Parameters of the best-fit multipolar magnetic field models of \hd.}
\label{tbl1}
\begin{tabular}{lccc}
\hline
\hline
Parameter & Model $A$ & Model $B$ & Model $C$ \\
\hline
$i$ (\degr)       & $81\pm15$    & $78\pm7$     & $64\pm21$    \\
$B_{\rm d}$ (kG)  & $3.1\pm1.8$  & $4.9\pm1.3$  & $5.5\pm2.2$  \\
$\beta$  (\degr)  & $64\pm12$    & $123\pm26$   & $84\pm16$    \\
$\gamma$ (\degr)  & $372\pm3$    & $367\pm16$   & $372\pm11$   \\ 
$B_{\rm q}$ (kG)  & $41.9\pm1.3$ & $-46.8\pm1.3$& $51.6\pm2.9$ \\
$\beta_1$ (\degr) &              & $43\pm6$     & $47\pm18$    \\
$\beta_2$ (\degr) &              & $=\beta_1$   & $114\pm18$   \\
$\gamma_1$ (\degr)&              & $81\pm4$     & $125\pm9$    \\
$\gamma_2$ (\degr)&              & $=\gamma_1$  & $216\pm10$   \\
$B_{\rm o}$ (kG)  & $-0.5\pm2.1$ &              &              \\
$\chi^2_\nu$      & 1.36         & 1.34         & 1.16         \\
\hline
\end{tabular}
\end{table}

\begin{figure*}[!t]
\figps{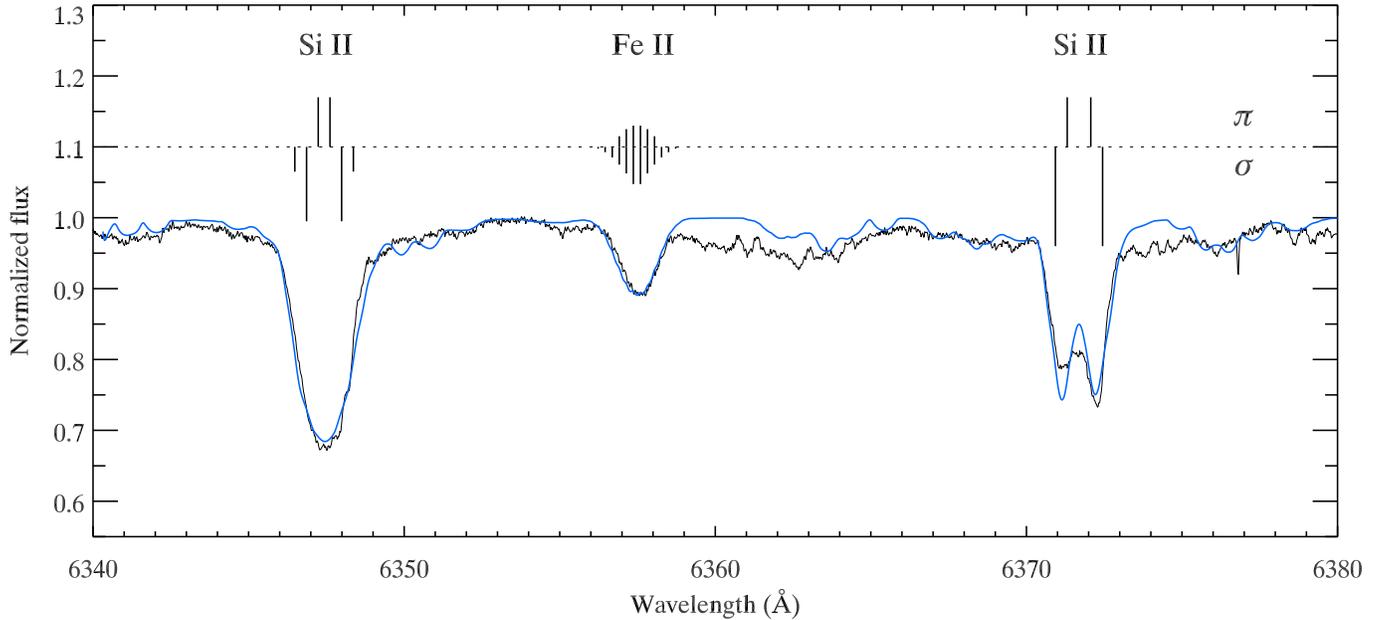}
\caption{Comparison of the observed (\textit{thin line}) and computed (\textit{thick line}) spectrum of \hd.
Synthetic spectrum is computed for $v_{\rm e}\sin i$\,=\,20~\kms\ and $\langle B\rangle$\,=\,29~kG using the {\sc synthmag} code.
Zeeman patterns are schematically illustrated for the \ion{Si}{ii} 6347.11, 6371.37~\AA\ and \ion{Fe}{ii} 6357.16~\AA\
lines. The length of vertical bars is proportional to the relative strength of the respective Zeeman components;
$\pi$ components are shown above the horizontal dotted line, and $\sigma_\pm$ components below this line.}
\label{fig3}
\end{figure*}

Parameters of all magnetic models were optimized using the least-squares procedure. Resulting best-fit
multipolar field characteristics are listed in Table~\ref{tbl1}. Theoretical phase curves of the four
magnetic observables are compared with available observations in Fig.~\ref{fig2}. The three magnetic
models yield reduced $\chi^2$ of 1.16--1.36, indicating acceptable fitting of the data. The general
superposition of dipole and non-linear quadrupole (model $C$) provides a marginally better description
of the magnetic measurements. It is encouraging to find consistency in some of the geometrical parameters
recovered in the fit with different multipolar parameterizations. All models favour large inclination
angle, $i$\,=\,64\degr--81\degr, and suggest similar strength and orientation for the dipolar
component. The strength of the dominant quadrupolar component ranges from
42 to 52~kG -- at least 9 times larger than the dipolar contribution. The projected rotational
velocity deduced from the crossover observations is in the interval 9--13~\kms, which is smaller than 
the spectroscopically determined \vs\ (see Sect.~\ref{split}). The stellar radius estimated using the
aforementioned range of possible stellar inclination and the rotation period of \hd\ is also
unrealistically small ($R\approx R_\odot$) for $v_{\rm e}\sin i$\,$\approx$\,10~\kms. On the other
hand, one finds $R\approx$\,1.8--3.0\,$R_\odot$ if $v_{\rm e}\sin i$\,$=$\,20--30~\kms\ is adopted.

The three magnetic models predict nearly identical rotational modulation of \bz\ and \bx. Expected
variation of \bq\ is marginally different, but corresponding theoretical curves cannot be distinguished
based on observations because of the large errors of the quadratic field measurements. Thus,
no unique multipolar field model can be derived for \hd\ with the available magnetic observations.
This situation is often encountered in the analysis of the disk-integrated magnetic observables
of magnetic CP stars when using some specific parameterization of magnetic geometry (Bagnulo et al. 
\cite{BLL02}) or when comparing results obtained with different multipolar parameterizations
(Kochukhov \cite{K04}). At the same time, one can see that
the mean surface field variation expected for the alternative magnetic models of \hd\ is substantially different,
especially for the rotation phase interval $\varphi$\,=0.6--1.0. Predicted \bs\ ranges from 22 to 31~kG.
Corresponding Zeeman splitting of the magnetically sensitive spectral lines should be comparable or larger
than the rotational Doppler broadening of the \hd\ spectrum (\vsv{28\pm5}\kms\ according to the estimate
by Mathys (\cite{M95b})). Thus, if one believes that the low-order multipolar field models provide an
adequate description of the non-dipolar field in \hd, there are good prospects for detecting Zeeman
split lines in high-resolution optical spectra of this star.

\section{Zeeman split lines in the spectrum of \hd}
\label{split}

High-resolution spectra of \hd\ were obtained with the UVES instrument of the ESO VLT on July 12, 2001
($JD=2452102.60$, phase 0.67 for the 4.4916~d rotation period).
Observations were carried out in the context of the 
\textit{UVES Paranal Observatory Project}\footnote{{\tt http://www.eso.org/uvespop/}} 
(ESO DDT program 266.D-5655, Bagnulo et al. \cite{BJL03}). The spectra were recorded in dichroic modes,
using the slit width of 0.5$^{\prime\prime}$. A spectral resolution of about 80000 and almost complete spectral coverage
of the wavelength interval from 3070 to 10400~\AA\ were achieved. The signal-to-noise ratio of the \hd\
spectrum is above 300 for the optical region. Reduction of the UVESpop observations is described by Bagnulo et
al. (\cite{BJL03}).

Inspection of the UVES spectra of \hd\ revealed a significant distortion in the profiles of many
spectral lines. In addition to the overall asymmetry of many features, some lines exhibit resolved 
absorption components which are considerably narrower than expected for the normal late B stellar
spectrum with a $v_{\rm e}\sin i$\,$\approx$30~\kms\ Doppler broadening. This may be a signature of the Zeeman
splitting in a very strong magnetic field or a manifestation of an inhomogeneous surface chemical
distribution. To separate these effects, I have investigated several groups of spectral lines, belonging
to the same ions, but with different sensitivity to magnetic field.

Remarkable proof of the presence of a very strong magnetic field in \hd\ comes from the analysis of the
red \ion{Si}{ii} doublet at $\lambda$ 6347 and 6371~\AA. These lines originate from the same lower
atomic level and have similar strength. Nevertheless, as evident from Fig.~\ref{fig3},
profile shape of the two \ion{Si}{ii} lines is very different. The 6371~\AA\ line shows an asymmetric
double  structure, whereas the \ion{Si}{ii} 6347~\AA\ lines exhibits no apparent anomaly. Obviously,
horizontal abundance spots cannot be responsible for such a drastically different shape of the two similar lines.
Consideration of the Zeeman effect allows one to understand the puzzling behaviour of the \ion{Si}{ii}
doublet. The  \ion{Si}{ii} 6371~\AA\ line has a relatively simple, nearly doublet, Zeeman splitting
pattern with two $\pi$ and two $\sigma$ components. On the other hand, the 6347~\AA\ line splits into  6
close Zeeman components, which are blended together and do not produce resolved structure in \hd. In
Fig.~\ref{fig3} I compare observations and theoretical spectrum calculated with the magnetic spectrum 
synthesis code {\sc synthmag} (Piskunov \cite{P99}). I have adopted \tev{12750}, \lggv{3.8}, 
$v_{\rm e}\sin i$\,=\,20~\kms, and \bs\,=\,29~kG. The Zeeman splitting of \ion{Si}{ii} 6371~\AA\ line and 
broadening of the \ion{Si}{ii} 6347~\AA\ and \ion{Fe}{ii} 6357~\AA\ lines are reproduced fairly well,
although a homogeneous magnetic field adopted in the {\sc synthmag} calculations is insufficient to model
relative strengths of the \ion{Si}{ii} 6371~\AA\ Zeeman components.

\begin{figure}[!t]
\figps{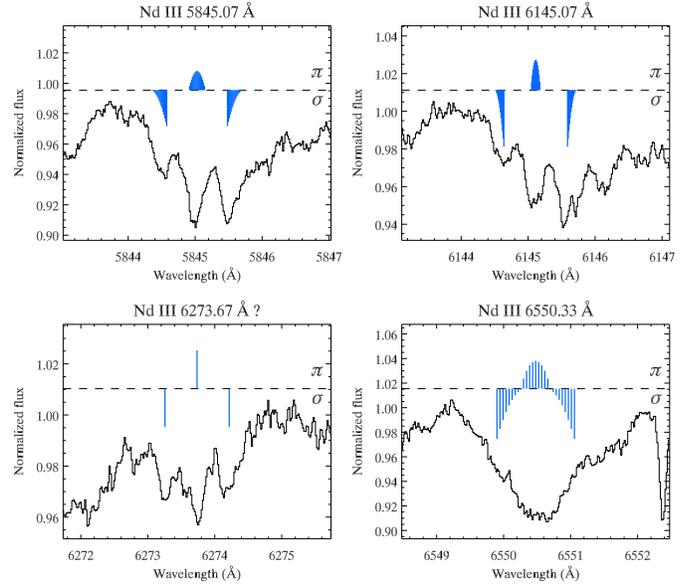}
\caption{Zeeman resolved \ion{Nd}{iii} lines in the spectrum of \hd. The four panels show lines with
different Zeeman splitting patterns. The strength and position of the $\pi$ and $\sigma$ Zeeman components
for $\langle B \rangle=29$~kG are illustrated as in Fig.~\ref{fig3}. The classical triplet pattern
with $z=0.9$ is adopted for the unclassified line \ion{Nd}{iii} 6273.67~\AA.}
\label{fig4}
\end{figure}

A number of doubly ionized Nd lines can be identified in the spectrum of \hd. Some of these features show
conspicuous narrow components in the triplet-like splitting patterns (Fig.~\ref{fig4}). Detailed
investigation of several lines confirms this to be a magnetic field effect. The \ion{Nd}{iii} lines with
significant Land\'e factors and quasi-triplet Zeeman splitting (e.g. 5845 and 6145~\AA) exhibit resolved
components. A well-defined triplet structure is also observed for an unidentified line at 6273.67~\AA.
According to the Nd laboratory line list compiled by H. Crosswhite, this line belongs to  the \ion{Nd}{iii}
spectrum (T. Ryabchikova, private communication). On the other hand, as expected from its small Land\'e 
factor and complex Zeeman pattern, the resonance \ion{Nd}{iii} 6550~\AA\ line shows no magnetic splitting. 
Resolved Zeeman components of the \ion{Nd}{iii} lines suggest  $v_{\rm e}\sin i$\,$\approx$\,10~\kms, which is
substantially smaller than the projected rotational velocity inferred from the \ion{Si}{ii} and other spectral
lines. This anomaly may be attributed to a high-contrast inhomogeneous horizontal distribution of Nd. In such
a situation Nd lines will sample a fraction of the visible stellar surface and, therefore, will be less
broadened by rotation.

To improve the accuracy of the magnetic field strength measurement, I have determined \bs\ from several
lines showing Zeeman components: \ion{Si}{ii} 6371, \ion{Fe}{ii} 5018, \ion{Ba}{ii} 6141, \ion{Nd}{iii}
5845 and 6145~\AA. Fitting superposition of Gaussian profiles to these five lines yields
\bs\,=\,$28.7\pm0.8$~kG (the quoted uncertainty is the standard error of the mean for the measurements obtained from
individual lines). The aforementioned spectrum synthesis modelling of the \ion{Si}{ii} doublet is in
agreement with this estimate of the surface field strength. 

The first field modulus measurement obtained for \hd\ in the present paper is compared with the multipolar
field model predictions in Fig.~\ref{fig2}. The maximum attained field strength is consistent with
observations for all three models. The non-axisymmetric superposition of the dipole and linear quadrupole
(model $B$, see Sect.~\ref{mag}) shows the best agreement with the observational estimate of \bs, whereas the
axisymmetric model $A$ is probably ruled out.

Magnetic spectrum synthesis with the {\sc synthmag} code was used to assess chemical abundance pattern of
\hd. Even when the very strong magnetic field is taken into account, one finds a large enhancement of the Si and
iron-peak element concentration with respect to the solar chemical composition: [Si/H]\,=\,$+0.8$,
[Fe/H]\,=\,$+1.4$, [Cr/H]\,=\,$+1.9$, and [Ti/H]\,=\,$+2.6$. In contrast, several light elements are
strongly underabundant: [He/H]\,$<$\,$-3.5$, [Ca/H]\,=\,$-2.2$, [Mg/H]\,=\,$-1.2$.

\section{Discussion}
\label{discus}

In this paper I have presented analysis of the magnetic field in the unique peculiar B-type star \hd.  This object
is one of the very few early-type stars with distinctively non-dipolar magnetic field geometry. Interpretation of the
previous magnetic measurements of \hd\ with the help of the low-order multipolar field models
suggests that the dominant contribution to the surface field in this star comes from the $>$\,40~kG quadrupolar
component. Mutipolar models predict the disk-averaged surface field strength in the range from 22 to
31~kG. My investigation of the high-resolution UVES spectra of \hd\ revealed resolved Zeeman splitting in several
spectral lines. The inferred field strength of 28.7~kG is in accord with the multipolar model predictions.

Unlike several other cases when a large quadratic field deduced for spotted magnetic CP
stars contradicts high-resolution line profile studies (see Kupka et al. (\cite{KRW96}) for $\alpha$~Cir and Kochukhov et
al. (\cite{KDP04}) for HR\,3831), the quadratic field measurements of \hd\ appear to be more reliable. 
This is likely to be a result of a comparatively strong influence of the magnetic field on the line profile shapes in
\hd. In stars with weaker fields effects of the horizontal chemical inhomogeneities often dominate the line profiles,
which distorts the quadratic field diagnostic (Kochukhov \cite{K04}) and sometimes leads to a spurious detection of  
strong magnetic fields.

The present study establishes \hd\ as an outstanding object with the second-largest field among CP and main
sequence stars in general. The only star with a stronger field is the well-known Babcock's star (HD\,215441).
The latter shows varying field modulus between 32 and 34~kG (Preston \cite{P69}). However, orientation of the rotation
axis and magnetic field geometry of HD\,215441 are not favourable for detailed analysis of the field topology
(Landstreet et al. \cite{LBB89}). In contrast, large inclination angle and unusual, very strong field of \hd\
open remarkable possibilities for the detailed magnetic field mapping (Piskunov \& Kochukhov \cite{PK02}) and for the analysis of
relation between chemical spots and magnetic structures. Outstanding chemical enrichment of the \hd\
atmosphere allows to use this star as a test ground for the application of the new generation 
model atmospheres of early-type stars, incorporating effects of anomalous chemical composition
and strong magnetic field (Shulyak et al. \cite{STR04}; Valyavin et al. \cite{VKP04}; 
Kochukhov et al. \cite{KKS05}).

The measurement of the field strength in \hd\ is the first case when we are able to impose a strong 
constraint on the fundamental field characteristic of an early-type star with a non-dipolar field topology. Good
agreement between \bs\ determination and the outcome of the multipolar field analysis supports the common (through
not as yet thoroughly verified) assumption that deviations from the basic dipolar field structure in the
early-type magnetic stars has the form of quadrupole or octupole. Justification of the mutipolar field modelling
framework hints that another star with the complex, double-wave variability of the longitudinal field -- HD\,37776
(Thompson \& Landstreet \cite{TL85}) -- also hosts a similarly large magnetic field, or may even possess the
strongest field of all known non-degenerate stars. Unfortunately, the rapid rotation of HD\,37776 precludes us
from detection of the Zeeman splitting in spectral lines and hence leaves no chance of direct confirmation
of the presence of an extraordinarily large field in this star.

\begin{acknowledgements}

The author is grateful to Dr. T. Ryabchikova for her valuable remarks and help with the \ion{Nd}{iii}
line identification. Resources provided by the electronic databases (VALD, SIMBAD, NASA's ADS)
are acknowledged. This work was supported by the grant from the \textit{Kungliga Fysiografiska
S\"allskapet i Lund}.

\end{acknowledgements}

\end{document}